\documentclass[draftclsnofoot, onecolumn, twoside, 12pt]{IEEEtran}

\usepackage{amsmath}
\usepackage{amssymb}
\usepackage{graphicx}
\usepackage{cite}
\usepackage{comment}
\usepackage{psfrag}
\usepackage{epsfig}
\usepackage{color}
\usepackage{graphicx}
\usepackage[nolist,printonlyused]{acronym}
\usepackage{srcltx}

\renewcommand{\QED}{\QEDopen}

\newtheorem{theorem}{Theorem}
\newtheorem{definition}{Definition}
\newtheorem{corollary}{Corollary}

\newtheorem{example}{Example}
\newtheorem{property}{Property}

\begin{document}

\setcounter{page}{0}


\title{Doubly-Generalized LDPC Codes: Stability Bound over the BEC}
\author{
    Enrico~Paolini,~\IEEEmembership{Member,~IEEE}, and \\
    Marc~Fossorier,~\IEEEmembership{Fellow,~IEEE}, and \\
    Marco~Chiani,~\IEEEmembership{Senior~Member,~IEEE}. \\
    \vspace{2cm} 
    \underline{Corresponding Address:}\\
    Marco Chiani\\
    DEIS, University of Bologna\\
    V.le Risorgimento 2\\
    40136 Bologna, ITALY\\
    \bigskip
    Tel: +39-051-2093084 \qquad Fax: +39-051-2093540\\
    e-mail: \texttt{marco.chiani@unibo.it}
\thanks{
        Enrico Paolini and Marco Chiani are with DEIS/WiLAB,
        University of Bologna,
        V.le Risorgimento 2,
        40136 Bologna, ITALY.
        E-mail: {\tt e.paolini@unibo.it}, {\tt mfossorier2@yahoo.com}, {\tt marco.chiani@unibo.it}.}
}
\markboth
    {submitted to IEEE Trans. Inform. Theory}
    {Doubly-Generalized LDPC Codes: Stability Bound over the BEC}
%
\maketitle

\thispagestyle{empty}

\newpage
\setcounter{page}{1}%
%
\begin{abstract}
The iterative decoding threshold of low-density parity-check
(LDPC) codes over the binary erasure channel (BEC) fulfills an
upper bound depending only on the variable and check nodes with
minimum distance 2. This bound is a consequence of the stability
condition, and is here referred to as stability bound. In this
paper, a stability bound over the BEC is developed for
doubly-generalized LDPC codes, where the variable and the check
nodes can be generic linear block codes, assuming maximum a
posteriori erasure correction at each node. It is proved that in
this generalized context as well the bound depends only on the
variable and check component codes with minimum distance 2. A
condition is also developed, namely the derivative matching
condition, under which the bound is achieved with equality.
\end{abstract} {\pagestyle{plain} \pagenumbering{arabic}}
%
%
\section{Introduction}\label{section:introduction}
LDPC codes \cite{gallager63:low-density} have been intensively
studied in the last decade due to their capability to approach the
Shannon limit under iterative belief-propagation decoding. An LDPC
code of length $N$ and dimension $K$ can be graphically
represented as a bipartite graph, known as Tanner graph, with $N$
variable nodes (VNs) and $M \geq N-K$ check nodes (CNs)
\cite{tanner81:recursive}. In the Tanner graph, the degree of
either a VN or a CN is defined as the number of edges connected to
it.
%
A degree-$n$ VN of an LDPC code can be interpreted as a length-$n$
repetition code, i.e., as a $(n,1)$ linear block code repeating
$n$ times its only information bit towards the check node decoder
(CND). Instead, a degree-$n$ CN of an LDPC code can be interpreted
as a length-$n$ single parity-check (SPC) code, i.e., as a
$(n,n-1)$ linear block code.

An extension of the concept of LDPC code is represented by
doubly-generalized LDPC (D-GLDPC) codes \cite{wang06:D-GLDPC},
where the VNs and the CNs are allowed to be generic $(n,k)$ linear
block codes instead of repetition and SPC codes, respectively. If
only the CND is generalized while all the VNs are repetition
codes, then the code is said a generalized LDPC (GLDPC) code, or a
Tanner code \cite{tanner81:recursive}.
%
%

\begin{figure}[t]
\begin{center}
\psfrag{n2edges}{\small{$n_2$ edges}} \psfrag{k1bits}{
\small{$k_1$ bits}} \psfrag{n2k2genCN}{$(n_2,k_2)$ generalized CN}
\psfrag{n2k2eqns}{\small{$n_2-k_2$ equations}} \psfrag{SPCCN}{SPC
CN} \psfrag{repVN}{rep. VN} \psfrag{n1k1genVN}{$(n_1,k_1)$
generalized VN}
\psfrag{brace}{$\underbrace{\phantom{---------------------------}}$}
\psfrag{encodedbits}{encoded bits} \psfrag{n1edges}{\small{$n_1$
edges}}
\includegraphics[width=7 cm, angle=270]{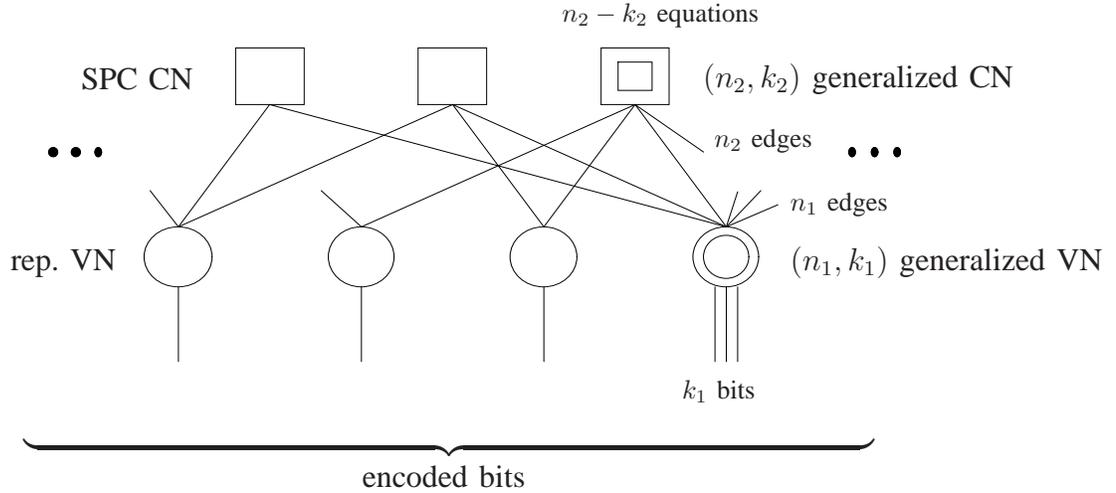}
\end{center}
\caption{Structure of a D-GLDPC code.} \label{fig:DGLDPC}
\end{figure}

In a D-GLDPC code the codes used as VNs and CNs are called
\emph{component codes}. In this work each component code is
supposed to be a linear block code having a minimum distance
$d_{\min} \geq 2$. The VNs and the CNs which are not repetition or
SPC codes, are referred to as \emph{generalized nodes}. The
corresponding code structure is depicted in Fig.~\ref{fig:DGLDPC}.
An $(n,k)$ generalized VN is characterized by $n$ connections
towards the CND; moreover, $k$ of the $N$ D-GLDPC encoded bits are
associated with the VN, and interpreted by the VN as its $k$
information bits. Then, the codeword length of a D-GLDPC code with
$N_V$ VNs is $N = \sum_{i=1}^{N_V} k_i$ ($k_i$ being the dimension
of the $i$-th VN). An $(n,k)$ generalized CN is characterized by
$n$ connections towards the variable node decoder (VND), and is
associated with $n-k$ independent parity check equations. Then,
the number of parity-check equations for a D-GLDPC code with $N_C$
CNs is $M = \sum_{i=1}^{N_C} (n_i - k_i)$ ($k_i$ and $n_i$ being
the dimension and length of the $i$-th check node, respectively).
For a description of a D-GLDPC code iterative decoder over the
AWGN channel and BEC we refer to \cite{wang06:D-GLDPC} and
\cite{paolini07:phd}, respectively.

For LDPC code ensembles, an important role is played by a theorem
known as the \emph{stability condition}
\cite{luby01:efficient,richardson01:design,urbanke:modern}. The
most important consequence of the stability condition is the
possibility to upper bound the asymptotic iterative decoding
threshold. If the communication channel is a BEC with erasure
probability $q$, the stability condition leads to the following
upper bound on the asymptotic threshold $q^*$ for the LDPC
ensemble:
\begin{align}\label{eq:stability-LDPC}
q^{*} \leq [\lambda'(0)\,\rho'(1)]^{-1}.
\end{align}
The inequality \eqref{eq:stability-LDPC} is referred to as
\emph{stability bound} in this paper\footnote{We use the
nomenclature stability bound as, even though
\eqref{eq:stability-LDPC} is sometimes referred to as the
stability condition, strictly speaking it is a consequence of it.
For more details we refer to \cite[Theorem 3.66]{urbanke:modern}
and the related discussion.}. In \eqref{eq:stability-LDPC},
$\lambda'(0) = \lambda_2$ is the fraction of edges connected to
the length-2 repetition VNs, while $\rho'(1)$ is the derivative
(computed at $x=1$) of the LDPC CNs degree distribution
$\rho(x)=\sum_{j \geq 2}\rho_j x^{j-1}$, where $\rho_j$ is the
fraction of edges connected to SPC CNs of length $j$. The bound
\eqref{eq:stability-LDPC} was first developed from density
evolution. Next we propose a simple graphical interpretation of
\eqref{eq:stability-LDPC} using extrinsic information transfer
(EXIT) charts \cite{ten-brink01:convergence}. Let us denote by
$I_A$ the average \emph{a priori} mutual information in input to
the VND or to the CND. Furthermore, let us denote by
$I_{E,V}(I_A,q)$ and $I_{E,C}(I_A)$ the average extrinsic
information for the VND and CND respectively (these functions are
usually referred to as \emph{EXIT functions}). Then,
\eqref{eq:stability-LDPC} is equivalent to the following
condition: for $q = q^*$, the derivative of the VND EXIT function
$I_{E,V}(I_A,q)$, with respect to $I_A$ and evaluated at $I_A=1$,
must be smaller than the derivative of the inverse CND EXIT
function $I_{E,C}^{-1}(I_A)$ evaluated at $I_A=1$. That is
\eqref{eq:stability-LDPC} is equivalent to requiring

\begin{align}\label{eq:stability-condition}
\frac{\partial I_{E,V}(I_A,q^*)}{\partial I_A}\,\Big|_{I_A=1} \leq
\frac{{\rm d}I_{E,C}^{-1}(I_A)}{{\rm d}I_A}\,\Big|_{I_A=1}.
\end{align}

There exist LDPC degree distributions achieving the bound
\eqref{eq:stability-LDPC} with equality, so that their threshold
over the BEC assumes the simple closed form $q^* =
[\lambda'(0)\rho'(1)]^{-1}$. For such LDPC distributions, the
first occurrence of a tangency point between the VND EXIT function
$I_{E,V}(I_A,q)$ and the inverse CND EXIT function
$I_{E,C}^{-1}(I_A)$ appears at $I_A=1$, i.e.

\begin{align}\label{eq:derivative-matching}
\left\{
\begin{array}{l}
I_{E,V}(1,q^*)=I_{E,C}^{-1}(1)\\
\frac{\partial I_{E,V}(I_A, \, q^*)}{\partial I_A} \,
\Big|_{I_A=1} = \frac{{\rm d}I_{E,C}^{-1}(I_A)}{{\rm d}I_A} \,
\Big|_{I_A=1}.
\end{array}\right.
\end{align}

\bigskip
For LDPC code ensembles characterized by VNs and CNs with degree
at least 2, the first equality is always satisfied as both terms
are equal to 1. This occurs also for D-GLDPC codes with all
variable and check component codes having a minimum distance
$d_{\min} \geq 2$ \cite{paolini07:doubly}, which is an assumption
of this paper. Then, only the second equality is considered in the
sequel, and is referred to as the \emph{derivative matching
condition}.

In this paper, the stability upper bound \eqref{eq:stability-LDPC}
and the derivative matching condition are extended to D-GLDPC
codes (and to GLDPC codes as a sub-case). Our derivations lead to
the conclusion that only the check and variable component codes
with minimum distance $d_{\min}=2$, including length-2 repetition
codes and SPC codes, contribute to the stability bound. We also
show that for D-GLDPC codes satisfying the derivative matching
condition the asymptotic threshold over the BEC can be expressed
by a simple formula.

The paper is organized as follows. Some definitions and the
notation used in the paper are introduced in Section
\ref{section:definitions}. In Section \ref{section:G-rank} the
possibility to reduce the rank of a linear block code generator
matrix by column elimination is discussed. Using these results, in
Section \ref{section:gldpc} and Section \ref{section:d-gldpc} the
stability bound is developed for GLDPC codes and for D-GLDPC
codes, respectively. Final remarks are given in Section
\ref{section:conclusion}.



\section{Definitions and Basic Notation}\label{section:definitions}
We assume as transmission channel a BEC with erasure probability
$q$. For a bipartite graph with random connections, the
\emph{extrinsic channel} (that is the channel over which the
messages are exchanged between the VND and the CND during the
iterative decoding process) is modelled as a second BEC with
erasure probability $p$ depending on the decoding iteration
\cite{ten-brink04:extrinsic}, where it is readily proved that $I_A
= 1-p$. Since we express both the VND and the CND EXIT functions
as functions of $p$ (and $q$ for the VND), their derivatives are
evaluated at $p=0$ (corresponding to $I_A=1$). In this case
\eqref{eq:stability-condition} becomes
\begin{align}\label{eq:p-perspective}
\frac{\partial I_{E,V}(p,q^*)}{\partial p}\,\Big|_{p=0} \geq
\frac{{\rm d}I_{E,C}(p)^{-1}}{{\rm d}p}\,\Big|_{p=0}.
\end{align}

Under the hypothesis of a random bipartite graph, the VND and CND
EXIT functions can be expressed as
\begin{align}\label{eq:exit-VND}
I_{E,V}(p,q) =
\sum_{i=1}^{\mathcal{I}_V}\lambda_i\,I_{E,V}^{(i)}(p,q)
\end{align}
and
\begin{align}\label{eq:exit-CND}
I_{E,C}(p) = \sum_{i=1}^{\mathcal{I}_C}\rho_i\,I_{E,C}^{(i)}(p),
\end{align}
respectively, where $\mathcal{I}_V$ and $\mathcal{I}_C$ are the
number of different VN and CN types, $I_{E,V}^{(i)}(p,q)$ and
$I_{E,C}^{(i)}(p)$ are the EXIT functions for the $i$-th VN type
and for the $i$-th CN type, respectively, and $\lambda_i$ and
$\rho_i$ are the fractions of edges towards the VNs of type $i$
and the CNs of type $i$, respectively.

For the sake of clarity, it is useful to isolate the contribution
of the repetition component codes in \eqref{eq:exit-VND} and the
contribution of the SPC component codes in \eqref{eq:exit-CND}, so
that
\begin{align}\label{eq:exit-VND-divide}
I_{E,V}(p,q) & = \sum_{j \geq 2}^{\rm (rep)}\lambda_j^{\rm (rep)
}\cdot(1-q\,p\,^{j-1}) + \sum_{i}^{\rm
(gen)}\lambda_i\,I_{E,V}^{(i)}(p,q) \notag \\
\, & = \sum_{j \geq 2}^{\rm (rep)}\lambda_j^{\rm (rep)} -
q\,\lambda_{\rm rep}(p) + \sum_{i}^{\rm
(gen)}\lambda_i\,I_{E,V}^{(i)}(p,q)
\end{align}
\begin{align}\label{eq:exit-CND-divide}
I_{E,C}(p) & = \sum_{j \geq 2}^{\rm (SPC)}\rho_j^{\rm (SPC)
}\cdot(1-p)^{j-1} +
\sum_{i}^{{\rm(gen)}}\rho_i\,I_{E,C}^{(i)}(p) \notag \\
\, & = \rho\,_{\rm{SPC}}(1-p) +
\sum_{i}^{{\rm(gen)}}\rho_i\,I_{E,C}^{(i)}(p).
\end{align}
In \eqref{eq:exit-VND-divide}, $j$ is the length of the generic
repetition VN, $\lambda_j^{\rm (rep)}$ is the fraction of edges
connected to the repetition VNs of length $j$, and $\lambda_{\rm
rep}(x) \triangleq \sum_{j \geq 2} \lambda_j^{\rm
(rep)}\,x^{j-1}$. We use in \eqref{eq:exit-VND-divide} the well
known EXIT function expression over the BEC for a $(j,1)$
repetition VN, i.e. $I_E(p,q) = 1 - q \, p\,^{j-1}$. The second
summation in \eqref{eq:exit-VND-divide} is over all the
generalized VN types. Analogously, in \eqref{eq:exit-CND-divide}
$j$ is the length of the generic SPC CN, $\rho_j^{\rm (SPC)}$ is
the fraction of edges towards the SPC CNs of length $j$,
$\rho_{\rm SPC}(x) \triangleq \sum_{j \geq 2} \rho_j^{\rm
(SPC)}\,x^{j-1}$, and we use the well-known EXIT function
expression over the BEC for a $(j\,,j-1)$ SPC CN, i.e. $I_E(p) =
(1-p)^{j-1}$.

The EXIT function of an $(n,k)$ generalized VN over the BEC, when
maximum a posteriori (MAP) erasure correction is performed at the
VN, can be expressed as
\begin{align}\label{eq:modified-IE-variable}
I_E(p,q) = 1 & - \frac{1}{n} \sum_{t=0}^{n-1}\,\sum_{z=0}^{k}
a_{t,z}\,p^t\,(1-p)^{n-t-1}\,q^z\,(1-q)^{k-z},
\end{align}
which can be readily obtained from \cite[eq.
36]{ten-brink04:extrinsic} with
$$a_{t,z} =
[(n-t) \, \tilde{e}_{n-t,k-z}-(t+1) \, \tilde{e}_{n-t-1,k-z}].$$
The parameter $\tilde{e}_{g,h}$ (with $g=0,\dots,n$ and
$h=0,\dots,k$) is known as the $(g,h)$-th un-normalized split
information function, defined as explained next. Considering a
representation $\mathbf{G}$ of the generator matrix for the
$(n,k)$ VN, and appending to it the $(k \times k)$ identity matrix
$\mathbf{I}_k$, $\tilde{e}_{g,h}$ is equal to the summation of the
ranks over all the possible submatrices obtained selecting $g$
columns out of $\mathbf{G}$ and $h$ columns out of $\mathbf{I}_k$.
We remark that the split information functions for a generalized
VN, and therefore its MAP EXIT function
\eqref{eq:modified-IE-variable}, depend on the chosen generator
matrix representation \cite{paolini07:doubly}. Then, the
performance of the overall D-GLDPC code depends on the code
representation used for the variable component codes. For the same
reason, two generalized VNs associated with the same code, but
with different generator matrices (i.e. different mappings between
information words and codewords) must be regarded in
\eqref{eq:exit-VND-divide} as VNs of different types.

The EXIT function of a generalized $(n,k)$ CN over the BEC, when
MAP decoding is performed at the CN, can be obtained by letting $q
\rightarrow 1$ in \eqref{eq:modified-IE-variable}. The obtained
expression, equivalent to \cite[eq. 40]{ten-brink04:extrinsic}, is
\begin{align}\label{eq:modified-IE-check}
I_E(p) = 1 - \frac{1}{n} \sum_{t=0}^{n-1} a_t \,
p^{t}(1-p)^{n-t-1},
\end{align}
with
$$a_t = (n-t)\,\tilde{e}_{n-t}-(t+1)\tilde{e}_{n-t-1}$$
For $g=0,\dots,n$, $\tilde{e}_g$ is known as the $g$-th
un-normalized information function of the $(n,k)$ code, a concept
first introduced in \cite{helleseth97:information}. It is defined
as the summation of the ranks over all the possible submatrices
obtained by selecting $g$ columns out of the generator matrix
$\mathbf{G}$. As opposed to the split information functions
$\tilde{e}_{g,h}$, the information functions $\tilde{e}_g$ are
independent of the code representation. Thus, different check
component code representations are associated with the same EXIT
function for the generalized CN. The performance of a GLDPC or
D-GLDPC code is then independent of the specific representation of
its generalized check component codes.

Let us suppose that a generic VN is a $(n, k)$ linear block code
$\mathcal{C}$, with generator matrix $\mathbf{G}$. We denote by
$\mathcal{C}'$ the $(n+k,k)$ linear block code generated by
$[\mathbf{G} | \mathbf{I}_{k}]$. The generic codeword of
$\mathcal{C}$ is denoted by $\mathbf{c}$, while the generic
codeword of $\mathcal{C}'$ by $\mathbf{c}'$. We have $\mathbf{c}'
= [\mathbf{c} | \mathbf{u}]$, where $\mathbf{c}$ and $\mathbf{u}$
must satisfy $\mathbf{c} = \mathbf{u} \, \mathbf{G}$: the code
$\mathcal{C}'$ then depends on the chosen generator matrix
representation for $\mathcal{C}$. It is readily shown that
$d_{\min}' \geq d_{\min} + 1$, where $d_{\min}$ and $d_{\min}'$
are the minimum distances of $\mathcal{C}$ and $\mathcal{C}'$,
respectively.

%
\section{Reducing a Generator Matrix Rank by Column Elimination}\label{section:G-rank}
For a given $(n,k)$ linear block code $\mathcal{C}$ and for a
given representation $\mathbf{G}$ of its generator matrix, we
denote by $\mathcal{S}_t$ a generic submatrix obtained by
selecting $t$ columns out of $\mathbf{G}$, and by
$\overline{\mathcal{S}}_t$ the submatrix composed of the $n-t$
remaining columns.

\bigskip
\begin{definition}
We say that $\overline{\mathcal{S}}_t$ covers a non-null codeword
$\mathbf{c} \in \mathcal{C}$ when there are no `1' positions of
$\mathbf{c}$ corresponding to columns belonging to
$\mathcal{S}_t$.
\end{definition}

\bigskip
\begin{example}\label{ex:cover}
Let us consider a $(7,3)$ simplex code with generator matrix
$$ \mathbf{G} =
\left[
\begin{array}{ccccccc}
1 & 0 & 0 & 1 & 1 & 0 & 1\\
0 & 1 & 0 & 1 & 0 & 1 & 1\\
0 & 0 & 1 & 0 & 1 & 1 & 1\\
\end{array}
\right]\,\, ,
$$

\noindent and let us denote by $\mathcal{S}_2$ the submatrix
composed of the last two columns of $\mathbf{G}$. Then, the only
non-null codeword covered by $\overline{\mathcal{S}}_2$ is
$[0,1,1,1,1,0,0]$.
\end{example}

\bigskip
The following theorem states that in order to reduce the rank of a
given generator matrix by column elimination, it is necessary and
sufficient that the removed pattern of columns covers at least one
non-null codeword.

\bigskip
\begin{theorem}\label{theorem:cover-codewords}
Let us consider an $(n,k)$ linear block code $\mathcal{C}$. For
any generator matrix representation, we have ${\rm rank}
(\mathcal{S}_t) < k$ if and only if $\overline{\mathcal{S}}_t$
covers at least one codeword.
\end{theorem}
\begin{proof}
[\emph{Sufficiency}] Suppose that $\overline{\mathcal{S}}_t$
covers a codeword $\mathbf{\hat{c}}$, and consider a
representation $\mathbf{\hat{G}}$ of the generator matrix where
$\mathbf{\hat{c}}$ is one of the rows. It follows that removing
from $\mathbf{\hat{G}}$ the $n-t$ columns associated with
$\overline{\mathcal{S}}_t$ reduces the rank because at least one
of the rows becomes an all-zero row, so that ${\rm rank}
(\mathcal{S}_t) < k$. Since any representation of the generator
matrix can be obtained from any other representation by row
summations only, and since row summations cannot modify the rank
of submatrices composed of generator matrix columns, we have ${\rm
rank} (\mathcal{S}_t) < k$ also for any representation other than
$\mathbf{\hat{G}}$.

[\emph{Necessity}] Conversely, let us suppose that ${\rm rank}
(\mathcal{S}_t) < k$ for a given generator matrix representation.
Using the same argument as for the sufficiency, we observe that
this inequality must be satisfied also for any other
representation of the generator matrix. As removing
$\overline{\mathcal{S}}_t$ from any generator matrix leads to a
$(k \times t)$ matrix with reduced rank, it must be possible to
obtain (from any generator matrix representation) a generator
matrix where one or more rows have only `0' in those positions
corresponding to $\mathcal{S}_t$. All these rows are non-null
codewords of $\mathcal{C}$ covered by $\overline{\mathcal{S}}_t$.
\end{proof}

\bigskip
\begin{corollary}\label{corollary:independent-set}
We have ${\rm rank} (\mathcal{S}_t) = k$ for all $\mathcal{S}_t$
if and only if $n - t < d_{\min}$.
\end{corollary}
\begin{proof}
[\emph{Sufficiency}] Let us suppose that ${\rm rank}
(\mathcal{S}_t) = k$ for all $\mathcal{S}_t$. By applying Theorem
\ref{theorem:cover-codewords} it follows that no submatrix
$\overline{\mathcal{S}}_t$ (composed of $n-t$ columns) can cover
any codeword. Then $n-t < d_{\min}$.

[\emph{Necessity}] Conversely, let us suppose that $n-t <
d_{\min}$. Then, no submatrix $\overline{\mathcal{S}}_t$ (composed
of $n-t$ columns) can cover any codeword. By applying Theorem
\ref{theorem:cover-codewords} we conclude that ${\rm rank}
(\mathcal{S}_t) = k$ for all $\mathcal{S}_t$.
\end{proof}

\bigskip
\begin{example} All the codewords of the $(7,3)$ simplex code of Example \ref{ex:cover}
have Hamming weight 4. As one of these codewords is
$[1,1,1,0,0,0,1]$, Theorem \ref{theorem:cover-codewords}
guarantees that if we remove the first three and the last column
from $\mathbf{G}$ given in Example \ref{ex:cover} (or from any
matrix obtained performing row summations on $\mathbf{G}$) we
obtain a $(3 \times 3)$ matrix with rank smaller than 3. On the
other hand, by Corollary \ref{corollary:independent-set} we know
that, even if we remove any set of three or less columns, the rank
of $\mathbf{G}$ remains unchanged.
\end{example}

\bigskip
In \cite{paolini07:doubly} the concept of \emph{independent set}
was introduced. Given a $(k \times n)$ rank-$r$ binary matrix, an
independent set of size $s$ is defined as any set of $s$ columns
such that removing these columns from the matrix leads to a $(k
\times (n-s))$ matrix with a rank smaller than $r$. By Theorem
\ref{theorem:cover-codewords} we now state that a necessary and
sufficient condition for a set of $s$ columns to be an independent
set of a $(k \times n)$ generator matrix is that the $s$ columns
cover at least one codeword. Moreover, by Corollary
\ref{corollary:independent-set} we recognize that any set of $s <
d_{\min}$ columns cannot form an independent set for the generator
matrix.
%

%
\section{Stability Bound and Derivative Matching for GLDPC Codes}\label{section:gldpc}
In GLDPC codes all the variable component codes are repetition
codes, which in \eqref{eq:exit-VND-divide} leads to $\sum_{j \geq
2}^{\rm (rep)}\lambda_j^{\rm (rep)}=1$. The EXIT function over the
BEC for the VND is then given by $I_{E,V}(p,q) = 1 - q
\lambda_{\rm rep}(x)$. It follows
\begin{align}\label{eq:derivative-variable-gldpc}
\frac{\partial I_{E,V}(p,q)}{\partial p} \Big|_{p=0} = -q \,
\lambda^{(\rm rep)}_2.
\end{align}

\noindent From \eqref{eq:exit-CND-divide}, the derivative of
$I_{E,C}(p)$ at $p=0$ is
\begin{align}\label{eq:exit-CND-derivative}
\frac{{\rm d} I_{E,C}(p)}{{\rm d} p} \,\Big|_{p=0} = -\rho'_{\rm
SPC}(1) + \sum_{i}^{{\rm(gen)}}\rho_i\,\frac{{\rm
d}I_{E,C}^{(i)}(p)}{{\rm d} p}\Big|_{p=0}.
\end{align}

\noindent In order to develop \eqref{eq:exit-CND-derivative} it is
necessary to explicit the derivative of each generalized CN type
EXIT function. This task can be performed by exploiting Corollary
\ref{corollary:independent-set}, as explained next.

Consider an $(n,k)$ generalized CN with EXIT function $I_E(p)$ in
the form \eqref{eq:modified-IE-check}. It is readily shown that
$$ \frac{{\rm d} I_E(p)}{{\rm d} p} \Big|_{p=0} = \frac{(n-1)a_0 -
a_1}{n} \, .
$$
We have $a_0 = 0$ if and only if the generalized CN has minimum
distance $d_{\min} \geq 2$. In fact, the generator matrix of the
check component code is full rank (rank $= k$) by definition, so
$\tilde{e}_n = k$. Furthermore, from Corollary
\ref{corollary:independent-set}, removing any single column from
the generator matrix does not reduce the rank if and only if
$d_{\min} \geq 2$, in which case we obtain $\tilde{e}_{n-k} =
n\,k$, so that $a_0 = n\,\tilde{e}_n - \tilde{e}_{n-1} = n\,k -
n\,k = 0$. As recalled in Section \ref{section:introduction}, the
hypothesis $d_{\min} \geq 2$ is always assumed in this paper.
Then, we can assume $a_0 = 0$.

If $d_{\min} \geq 2$ for the CN we obtain
$$ \frac{{\rm d} I_E(p,q)}{{\rm d} p} \,\Big|_{p=0} = - \frac{a_1}{n},
$$
where $a_1 = (n-1)\tilde{e}_{n-1} - 2\,\tilde{e}_{n-2} =
k\,n\,(n-1)- 2\,\tilde{e}_{n-2}$. By applying again Corollary
\ref{corollary:independent-set}, we obtain
\begin{align}\label{eq:a1-expression}
a_1 \left\{ \begin{array}{l}
= 0 \quad {\rm if } \,\,d_{\min} \geq 3\\
> 0 \quad {\rm if } \,\,d_{\min} = 2\,. \end{array} \right.
\end{align}
If the CN exhibits a minimum distance $d_{\min} \geq 3$, then
removing any pair of columns from the generator matrix does not
affect the rank. In this case $2\,\tilde{e}_{n-2} = 2 \, k\,{n
\choose 2} = k\,n\,(n-1)$, hence $a_1 = 0$.

According to these results, the only generalized CNs that
contribute to the summation in the second term of
\eqref{eq:exit-CND-derivative} are those characterized by
$d_{\min} = 2$. By recalling that all the SPC codes have minimum
distance 2, we conclude that \eqref{eq:exit-CND-derivative} only
depends on the check component codes with $d_{\min} = 2$. The
derivative at $p=0$ of the CND EXIT function can be then expressed
as

\begin{align}\label{eq:exit-CND-deriv-d2}
\frac{{\rm d} I_{E,C}(p)}{{\rm d} p} \,\Big|_{p=0} \, & =
-\rho'_{\rm SPC}(1) - \sum_{i}^{[2]}\rho_i\,\frac{k_i n_i
(n_i-1)-2\,\tilde{e}_{n_i-2}}{n_i} \notag \\
\, & = -\rho'_{\rm SPC}(1) - \sum_{i}^{[2]} \rho_i \frac{2
\Delta_{n-2}^{(i)} }{n_i}\,
\end{align}
where the notation $\sum^{[2]}$ is adopted to indicate the
summation over those generalized CN types with minimum distance 2.
In \eqref{eq:exit-CND-deriv-d2}, we have denoted by
$\Delta_{n-2}^{(i)}$ the expression $k_i n_i (n_i-1)/2 -
\tilde{e}_{n_i-2}$, that does not depend on the chosen
representation for the $i$-th generalized CN type.

The next theorem states that $\Delta_{n-2}^{(i)}$ is equal to the
multiplicity $A^{(i)}_2$ of the weight-2 codewords for the CNs of
type $i$.

\bigskip
\begin{theorem}\label{theo:A2}
For any linear block check component code with minimum distance
$d_{\min}=2$, the parameter $\Delta_{n-2}$ equals the multiplicity
$A_2$ of the CN codewords with Hamming weight 2, i.e.
$$
\Delta_{n-2} = A_2\, .
$$
\end{theorem}
\begin{proof}
Let $\mathcal{S}_{n-2}$ be the generic $(k \times (n-2) )$ matrix
obtained by removing 2 columns from (any representation of) the CN
generator matrix. By Corollary \ref{corollary:independent-set} we
have that either ${\rm rank}(\mathcal{S}_{n-2}) = k$ or ${\rm
rank}(\mathcal{S}_{n-2}) = k-1$: considering a CN with
$d_{\min}=2$, removing any single column cannot reduce the rank so
that removing two columns can reduce the rank at most by one.

We have
\begin{align*}
\Delta_{n-2} & = \frac{k n (n-1)}{2} - \tilde{e}_{n-2} \\
\, & = \sum_{\mathcal{S}_{n-2}} k - \sum_{\mathcal{S}_{n-2}}
{\textrm{rank}} \left( \mathcal{S}_{n-2} \right) \\
\, & = \sum_{\mathcal{S}_{n-2}} \left( k - {\rm rank} \left(
\mathcal{S}_{n-2} \right) \right),
\end{align*}
where we know that each term in the summation is either equal to 0
or to 1. By Theorem~\ref{theorem:cover-codewords} any such term is
equal to 1 if and only if $\overline{\mathcal{S}}_{n-2}$ covers a
(necessarily weight-2) codeword.
\end{proof}

\bigskip
The derivative at $p=0$ of the inverse CND EXIT function
$I_{E,C}^{-1}(p)$ is given by $1 / {\textrm{d}} I_E (p) /
{\textrm{d}} p |_{p=0}$. Combining
\eqref{eq:derivative-variable-gldpc}, \eqref{eq:exit-CND-deriv-d2}
and Theorem~\ref{theo:A2}, for GLDPC codes,
\eqref{eq:p-perspective} becomes

\begin{align}\label{eq:stability-GLDPC}
q^* \leq \Big[ \lambda^{(\rm rep)}_2 \Big( \rho'_{\rm SPC}(1) +
\sum_{i}^{[2]} \rho_i \frac{2 A_2^{(i)}}{n_i}\ \Big) \Big]^{-1} \,
.
\end{align}



\noindent We can further simplify \eqref{eq:stability-GLDPC} by
noting that
\begin{align*}
\rho'_{\rm SPC}(1) & = \sum_j^{\rm (SPC)} \rho_j \, (j-1) \\
%
\, & = \sum_j^{\rm (SPC)} \rho_j \, \frac{2 \, A_2^{(j)}}{n_j} \,
,
\end{align*}
as for a SPC CN $n_j = j$ and $A_2^{(j)} = {j \choose 2} = j \,
(j-1) /2$. Hence, \eqref{eq:stability-GLDPC} can be written in the
more compact form
\begin{align}\label{eq:stability-dgldpc2}
q^* & \leq \left[\lambda^{\rm (rep)}_2 \,
\sum_{i}^{[2]}\frac{2\rho_i}{n_i}\,A_2^{(i)} \right]^{-1} \notag
\\
\, & = \left[\lambda^{\rm (rep)}_2 \, C \right]^{-1}
\end{align}
where
$$
C = \sum_i^{[2]} \rho_i \, C_i \qquad \textrm{with} \qquad C_i =
\frac{2 A^{(i)}_2}{n_i} \, ,
$$
and where now $\sum^{[2]}$ indicates the summation over all the
$d_{\min} = 2$ check component codes, both SPC and generalized.

For GLDPC codes satisfying the derivative matching condition
\eqref{eq:derivative-matching} (the first occurrence of a tangency
point between $I_{E,V}(p,q)$ and $I_{E,C}^{-1}(p)$ appears at
$p=0$), the threshold assumes the simple closed form
$ q^* = [\lambda^{\rm (rep)}_2 \, C ]^{-1}$.
If only generalized CNs with $d_{\min} \geq 3$ are used, then
\eqref{eq:stability-GLDPC} becomes $q^* \leq [ \lambda^{(\rm
rep)}_2 \, \rho'_{\rm SPC}(1) ]^{-1}$. If the derivative matching
condition is fulfilled in this case, we obtain $\, q^* = [
\lambda^{(\rm rep)}_2 \, \rho'_{\rm SPC}(1) ]^{-1}$.
%

%
\section{Stability Bound and Derivative Matching for D-GLDPC Codes}\label{section:d-gldpc}
The derivative at $p=0$ of the CND EXIT function of D-GLDPC codes
is the same as for GLDPC codes, that is ${\textrm{d}} I_E^{-1}(p)
/ {\textrm{d}} p |_{p=0} = - 1/C$. The partial derivative of the
VND EXIT function with respect to $p$ and evaluated at $p=0$, is
developed next.

It follows from \eqref{eq:exit-VND-divide} that
\begin{align}\label{eq:derivative-EXIT-variable-div}
\frac{\partial I_{E,V}(p,q)}{\partial p} \Big|_{p=0} =  -
q\,\lambda^{\rm (rep)}_2 + \sum_{i}^{\rm (gen)}\lambda_i\,
\frac{\partial I_{E,V}^{(i)}(p,q)}{\partial p} \Big|_{p=0}.
\end{align}
In order to develop the summation over the generalized VN types in
the second part of \eqref{eq:derivative-EXIT-variable-div}, we
have to explicit the partial derivative respect to $p$ of each
generalized VN type EXIT function, evaluated at $p=0$. To this
end, let us consider an $(n,k)$ generalized VN whose EXIT function
is given by \eqref{eq:modified-IE-variable}. After defining
$$f(p) = \sum_{t=0}^{n-1}\,a_{t,z}\,p\,^t\,(1-p)^{n-1-t}$$
we have
\begin{align}\label{eq:DGLDPC-partial}
\frac{\partial I_E(p,q)}{\partial p} \Big|_{p=0} & = -\frac{1}{n}
\, \sum_{z=0}^{k}\, \left(\frac{{\rm d}f(p)}{{\rm d}
p}\Big|_{p=0}\right) q^z\,(1-q)^{k-z} \notag \\
\, & = \sum_{z=0}^{k} \, \frac{(n-1)a_{0,z} - a_{1,z}}{n}\,\,
q^z\,(1-q)^{k-z},
\end{align}
as it is readily shown that that ${\rm d}f(p)/{\rm d}p\,|_{p=0} =
-(n-1)a_{0,z} + a_{1,z}$. The expression \eqref{eq:DGLDPC-partial}
can be further developed by invoking Corollary
\ref{corollary:independent-set}. Since any variable component code
has minimum distance $d_{\min} \geq 2$ by hypothesis, removing any
single column from the generator matrix $\mathbf{G}$ of the
variable component code cannot reduce the rank of $\mathbf{G}$. It
follows
\begin{align*}
a_{0,z} & = n\, \tilde{e}_{n,k-z}-\tilde{e}_{n-1,k-z}\\
\, & = k\,n\,{k \choose k-z} - k\,n\,{k \choose k-z} \\
\, & = 0,
\end{align*}
thus leading to
$$ \frac{\partial I_E(p,q)}{\partial p} \Big|_{p=0} = -\,
\sum_{z=0}^{k} \, \frac{a_{1,z}}{n} \,\, q^z\,(1-q)^{k-z} \, . $$

\noindent Corollary \ref{corollary:independent-set} can be invoked
again in order to show that
\begin{align}\label{eq:a1z-expression}
a_{1,z} \left\{ \begin{array}{l}
= 0 \quad \forall \, z \quad {\rm if } \,\,d_{\min} \geq 3\\
> 0 \quad \forall \, z \quad {\rm if } \,\,d_{\min} = 2,
\end{array} \right.
\end{align}
where $d_{\min}$ is the variable component code minimum distance.
In fact, under the hypothesis $d_{\min} \geq 3$, removing any
single column or any pair of columns from (any representation of)
$\mathbf{G}$ cannot reduce its rank. Under this hypothesis
\begin{align*}
a_{1,z} & = (n-1)\,\tilde{e}_{n-1,k-z} - 2\,\tilde{e}_{n-2,k-z}\\
\, & = k\,n\,(n-1)\,{k \choose k-z} - 2\,k\,{n \choose n-2}\,{k
\choose k-z}\\
\, & = 0.
\end{align*}

Hence, the only generalized variable component codes contributing
to \eqref{eq:derivative-EXIT-variable-div} are those with minimum
distance $d_{\min}=2$. This is coherent with the fact that, among
the repetition component codes, only those with $d_{\min}=2$ (i.e.
the length-2 repetition codes) give a non-null contribution to
\eqref{eq:derivative-EXIT-variable-div}.

Then, \eqref{eq:derivative-EXIT-variable-div} can be developed as
\begin{align}\label{eq:derivative-EXIT-variable-d2}
\frac{\partial I_{E,V}(p,q)}{\partial p} \Big|_{p=0} & =  - \,
q\,\lambda^{\rm (rep)}_2 -\sum_{i}^{[2]} \, \lambda_i \,
\sum_{z=0}^{k_i} \, \frac{k_i\,n_i(n_i-1){k_i \choose k_i-z}-2\,\tilde{e}_{n_i-2,k_i-z}}{n_i} \,\, q^z\,(1-q)^{k_i-z} \notag \\
\, & = - \, q\,\lambda^{\rm (rep)}_2 -\sum_{i}^{[2]} \, \lambda_i
\, \sum_{z=0}^{k_i} \, \frac{2 \Delta_{n-2,k-z}^{(i)}}{n_i} \,\,
q^z\,(1-q)^{k_i-z}.
\end{align}
In the previous expression the symbol $\sum^{[2]}$ indicates the
summation over those generalized VN types with minimum distance 2.
Moreover, $\Delta_{n-2,k-z}^{(i)}$ is defined as $\frac{k_i n_i
(n_i - 1)}{2} {k_i \choose k_i-z} - \tilde{e}_{n_i-z,k_i-z}$. As
opposed to $\Delta_{n-2}^{(i)}$ in \eqref{eq:exit-CND-deriv-d2},
$\Delta_{n-2,k-z}^{(i)}$ in \eqref{eq:derivative-EXIT-variable-d2}
depends on the component code representation.

Using \eqref{eq:derivative-EXIT-variable-d2} we can express
\eqref{eq:p-perspective} for a D-GLDPC code as
\begin{align}\label{eq:stability-dgldpc}
\, & q^*\,\lambda^{\rm (rep)}_2 +\sum_{i}^{[2]} \, \lambda_i \,
\sum_{z=0}^{k_i} \, \frac{2\,\Delta_{n-2,k-z}^{(i)}}{n_i} \,\,
(q^*)^z\,(1-q^*)^{k_i-z} \leq \frac{1}{C}.
\end{align}
In the reminder of this section, we prove that
\eqref{eq:stability-dgldpc} can be written as an explicit upper
bound to the decoding threshold $q^*$. We start by proving the
following theorem.

\bigskip
\begin{theorem}\label{theorem:sum-z:0...ki}
Let us consider an $(n,k)$ linear block variable component code
with minimum distance $d_{\min} = 2$. We have
\begin{align}\label{eq:sum-z:0...ki}
\sum_{z=0}^{k} \, \frac{2\,\Delta_{n-2,k-z}}{n} \,\,
(q^*)^z\,(1-q^*)^{k-z} = \sum_{u=1}^{k} \frac{2 \, A_{2,u}}{n}
\,\, (q^*)^u,
\end{align}
where $A_{2,u}$ is the number of the VN weight-2 codewords
generated by weight-$u$ information words.
\end{theorem}
\begin{proof}
Let $\mathcal{C}$ be the $(n,k)$ variable component code and let
$\mathbf{G}$ be the chosen generator matrix for $\mathcal{C}$.
Moreover, let $\mathcal{S}_{n-2,k-z}$ be the generic $(k \times
(n-2+k-z) )$ matrix obtained by selecting $n - 2$ columns in
$\mathbf{G}$ and $k-z$ columns in the $(k \times k)$ identity
matrix.

Let us apply Theorem \ref{theorem:cover-codewords} to the code
$\mathcal{C}'$ introduced at the end of Section
\ref{section:definitions}. Each codeword $\mathbf{c}' \in
\mathcal{C}'$ is composed of the concatenation of a codeword
$\mathbf{c} \in \mathcal{C}$ with one of the possible $2^k$
sequences of $k$ bits (where by the linearity of $\mathcal{C}$ the
all-zero length-$k$ sequence is always concatenated with the
all-zero codeword of $\mathcal{C}$). Combining this observation
with Theorem \ref{theorem:cover-codewords} and introducing the
notation $\mathcal{S}_{n-2,k-z} = [\mathcal{S}^{\mathbf{G}}_{n-2}
| \mathcal{S}^{\mathbf{I}}_{k-z}]$ we observe that a necessary
(though not sufficient) condition for having ${\rm
rank}(\mathcal{S}_{n-2,k-z}) < k$ is that
$\overline{\mathcal{S}}^{\mathbf{G}}_{n-2}$ covers a weight-2
codeword of $\mathcal{C}$.

Next, we develop $\Delta_{n-2,k-z}$ as
\begin{align*}
\Delta_{n-2,k-z} & = \frac{k n (n-1)}{2} {k \choose k-z} -
\tilde{e}_{n-2,k-z} \\
\, & = \sum_{\mathcal{S}_{n-2,k-z}} k -
\sum_{\mathcal{S}_{n-2,k-z}} {\rm rank}(\mathcal{S}_{n-2,k-z}) \\
\, & = \sum_{\mathcal{S}_{n-2,k-z}} \left(
k - {\rm rank}(\mathcal{S}_{n-2,k-z}) \right) \\
\, & = \sum_{\mathcal{S}^{\mathbf{G}}_{n-2}}
\sum_{\mathcal{S}^{\mathbf{I}}_{k-z}} \left( k - {\rm rank} \left(
[\mathcal{S}^{\mathbf{G}}_{n-2} |
\mathcal{S}^{\mathbf{I}}_{k-z}] \right) \right) \\
\, & = \sum_{\mathbf{c}}^{[2]}
\sum_{\mathcal{S}^{\mathbf{I}}_{k-z}} \left( k - {\rm rank} \left(
[\mathcal{S}^{\mathbf{G}}_{n-2} | \mathcal{S}^{\mathbf{I}}_{k-z}]
\right) \right),
\end{align*}

\noindent where $\sum_{\mathbf{c}}^{[2]}$ is used to indicate the
summation over those $\mathcal{S}^{\mathbf{G}}_{n-2}$ such that
$\overline{\mathcal{S}}^{\,\mathbf{G}}_{n-2}$ covers a weight-2
codeword of $\mathcal{C}$. Then, we can write

\begin{align}\label{eq:stability-dgldpc3}
\sum_{z=0}^{k} \, \frac{2\,\Delta_{n-2,k-z}}{n} & \,\,
(q^*)^z\,(1-q^*)^{k-z} \notag \\
\, & = \frac{2}{n} \sum_{z=0}^{k}\, \sum_{\mathbf{c}}^{[2]}
\sum_{\mathcal{S}^{\mathbf{I}}_{k-z}} \left( k - {\rm rank} \left(
[\mathcal{S}^{\mathbf{G}}_{n-2} | \mathcal{S}^{\mathbf{I}}_{k-z}]
\right) \right)\,
(q^*)^z\,(1-q^*)^{k-z} \notag \\
\, & = \frac{2}{n} \sum_{\mathbf{c}}^{[2]}\, \sum_{z=0}^{k}\,
\sum_{\mathcal{S}^{\mathbf{I}}_{k-z}} \left( k - {\rm rank} \left(
[\mathcal{S}^{\mathbf{G}}_{n-2} | \mathcal{S}^{\mathbf{I}}_{k-z}]
\right) \right)\, (q^*)^z\,(1-q^*)^{k-z} .
\end{align}

By hypothesis there are no VNs with minimum distance 1. Then, for
a given weight-2 codeword $\mathbf{c} \in \mathcal{C}$, any
submatrix $\mathcal{S}_{n-2,k-z}$ is such that
$\overline{\mathcal{S}}_{n-2,k-z}$ can cover at most one codeword
of $\mathcal{C}'$, i.e. the codeword $[\mathbf{c} |
\mathbf{u}_{\mathbf{c}}]$ subject to $\mathbf{c} =
\mathbf{u}_{\mathbf{c}} \, \mathbf{G}$.
If we denote by $w_H(\mathbf{u}_{\mathbf{c}})$ the Hamming weight
of $\mathbf{u}_{\mathbf{c}}$, for each weight-2 codeword
$\mathbf{c} \in \mathcal{C}$ the summation over $z$ in
\eqref{eq:stability-dgldpc3} can always start from
$w_H(\mathbf{u}_{\mathbf{c}})$. In fact, for $z = 0, \dots,
w_H(\mathbf{u}_{\mathbf{c}})-1$ it is not possible for
$\overline{\mathcal{S}}_{n-2,k-z}$ to cover the codeword
$[\mathbf{c} | \mathbf{u}_{\mathbf{c}}]$, so that $k - {\rm rank}
\left( [\mathcal{S}^{\mathbf{G}}_{n-2} |
\mathcal{S}^{\mathbf{I}}_{k-z}] \right) = 0$. That allows writing
the second member in \eqref{eq:stability-dgldpc3} as
\begin{align}\label{eq:stability-dgldpc4}
\frac{2}{n} \sum_{\mathbf{c}}^{[2]}\,
\sum_{z=w_H(\mathbf{u}_{\mathbf{c}})}^{k}\,
\sum_{\mathcal{S}^{\mathbf{I}}_{k-z}} \left( k - {\rm rank} \left(
[\mathcal{S}^{\mathbf{G}}_{n-2} | \mathcal{S}^{\mathbf{I}}_{k-z}]
\right) \right)\, (q^*)^z\,(1-q^*)^{k-z} .
\end{align}

For given $z \geq w_H \left( \mathbf{u}_{\mathbf{c}} \right)$, the
codeword $[\mathbf{c} | \mathbf{u}_{\mathbf{c}}]$ is covered by
exactly ${k-w_H \left(\mathbf{u}_{\mathbf{c}} \right) \choose
z-w_H(\mathbf{u}_{\mathbf{c}})}$ matrices
$\overline{\mathcal{S}}_{n-2,k-z}$. Hence, there are exactly
${k-w_H(\mathbf{u}_{\mathbf{c}}) \choose
z-w_H(\mathbf{u}_{\mathbf{c}})}$ non-null terms in

$$
\sum_{\mathcal{S}^{\mathbf{I}}_{k-z}} \left( k - {\rm
rank}(\mathcal{S}^{\mathbf{G}}_{n-2} |
\mathcal{S}^{\mathbf{I}}_{k-z}) \right) \, (q^*)^z\,(1-q^*)^{k-z}
\,  .
$$

\noindent Deleting from $\mathbf{G}$ two columns corresponding to
a weight-2 codeword of $\mathcal{C}$ reduces the rank of this
matrix by one, leading to a rank $k-1$. In fact, considering the
VN minimum distance $d_{\min}=2$, removing the first column cannot
reduce the rank (Corollary \ref{corollary:independent-set}) and
removing the second column reduces the rank (Theorem
\ref{theorem:cover-codewords}) necessarily by one. We can then
conclude that each of the ${k-w_H(\mathbf{u}_{\mathbf{c}}) \choose
z-w_H(\mathbf{u}_{\mathbf{c}})}$ non-null terms in the summation
$\sum_{\mathcal{S}^{\mathbf{I}}_{k-z}} \left( k - {\rm rank}\left(
[ \mathcal{S}^{\mathbf{G}}_{n-2} | \mathcal{S}^{\mathbf{I}}_{k-z}]
\right) \right)$ is equal to one, independently of $z$. Then we
can further develop \eqref{eq:stability-dgldpc4} as

\begin{align}\label{eq:stability-dgldpc5}
\frac{2}{n} \sum_{\mathbf{c}}^{[2]}\,
\sum_{z=w_H(\mathbf{u}_{\mathbf{c}})}^{k}\,
{k-w_H(\mathbf{u}_{\mathbf{c}}) \choose z-w_H \left(
\mathbf{u}_{\mathbf{c}} \right) }\, (q^*)^z\,(1-q^*)^{k-z} .
\end{align}

We next observe that those weight-2 codewords $\mathbf{c} \in
\mathcal{C}$ associated with the same
$w_H(\mathbf{u}_{\mathbf{c}})$ (i.e. generated by information
words having the same weight) produce the same contribution in
\eqref{eq:stability-dgldpc5}, since only the Hamming weight of the
information words $\mathbf{u}_{\mathbf{c}}$ matters.
This observation allows us to write \eqref{eq:stability-dgldpc5}
as
\begin{align}\label{eq:stability-dgldpc6}
\sum_{u=1}^{k} \, \frac{2\,A_{2,u}}{n} \sum_{z=u}^{k} \, {k-u
\choose z-u}\, (q^*)^z\,(1-q^*)^{k-z},
\end{align}
\noindent where $A_{2,u}$ is the number of weight-2 codewords
$\mathbf{c} \in \mathcal{C}$ such that
$w_H(\mathbf{u}_{\mathbf{c}}) = u$. In general, $A_{2,u}$ depends
on the variable component code representation.
By noting that
$$\sum_{z=u}^{k} \, {k-u \choose z-u}\,
(q^*)^z\,(1-q^*)^{k-z} = (q^*)^u \, ,$$
we finally obtain \eqref{eq:sum-z:0...ki}.
\end{proof}

\bigskip
Theorem \ref{theorem:sum-z:0...ki} allows us to write the first
member of \eqref{eq:stability-dgldpc} as
$$
q^*\,\lambda^{\rm (rep)}_2 +\sum_{i}^{[2]} \, \lambda_i \,
\sum_{u=1}^{k_i} \, \frac{2\,A^{(i)}_{2,u}}{n_i} (q^*)^u \, .$$
The length-2 repetition VNs can be embedded into the summation
over the generalized VN types with minimum distance 2. In fact,
the only weight-2 codeword of a length-2 repetition VN is
$\mathbf{c} = [1, 1]$, which is generated by a weight-1
information word. Then, for a length-2 repetition VN we have
\begin{align*}
\lambda_2^{\textrm{(rep)}}\, \sum_{u=1}^{k} \,
\frac{2\,A_{2,u}}{n} (q^*)^u = \lambda_2^{\textrm{(rep)}}\,\, q^*
\, .
\end{align*}

\noindent Hence, \eqref{eq:stability-dgldpc} can be put into the
more compact form
\begin{align}\label{eq:pre-final-stability-D-GLDPC}
\sum_{i}^{[2]} \, \lambda_i \, \sum_{u=1}^{k_i} \,
\frac{2\,A^{(i)}_{2,u}}{n_i} (q^*)^u & \leq \frac{1}{C} \, ,
\end{align}

\noindent where now the summation $\sum^{[2]}$ is over all the VN
types with minimum distance 2, both repetition and generalized.

The first part of \eqref{eq:pre-final-stability-D-GLDPC} is a real
polynomial $P(\cdot)$ in the variable $q^*$. This polynomial can
be written as $P(x) = \sum_i^{[2]} \, \lambda_i \, P_i(x)$, where
$P_i(\cdot)$ is a degree-$k_i$ real polynomial associated with the
$d_{\min}=2$ type-$i$ VNs. Each $P_i(\cdot)$ is a monotonically
increasing function (since all its coefficients are positive).
Consequently, $P(\cdot)$ is a monotonically increasing function
and its inverse $P^{-1}(\cdot)$ exists. We have then proved the
following theorem, which is the main contribution of this paper.

\bigskip
\begin{theorem}[Stability bound over the BEC for D-GLDPC
codes]\label{theorem:DGLDPC-bound}
The asymptotic threshold $q^*$ of a D-GLDPC code ensemble over the
BEC, assuming MAP erasure correction at each component code,
fulfills
\begin{align}\label{eq:general-stability-cond}
q^* \leq P^{-1} \left( \frac{1}{C} \right),
\end{align}
\noindent where
$$
P(x) = \sum_i^{[2]} \lambda_i \, P_i (x) \qquad {\rm with} \qquad
P_i(x) = \sum_{u=1}^{k_i} \, \frac{2\,A^{(i)}_{2,u}}{n_i}\,\, x^u
$$
\noindent and
$$
C = \sum_i^{[2]} \rho_i \, C_i \qquad {\rm with} \qquad C_i =
\frac{2\,A_2^{(i)}}{n_i} \, .
$$
\hspace*{\fill}~\QED\par\endtrivlist\unskip
\end{theorem}

\bigskip For an LDPC
code ensemble \eqref{eq:general-stability-cond} returns $q^* \leq
[\lambda'(0)\, \rho'(1)]^{-1}$, i.e. the well-known stability
bound for LDPC codes. 
%
\medskip
\begin{property}\label{prop:1}
For a length-2 repetition VN (i.e. for a conventional LDPC
degree-2 VN) we have $P_i(x) = x$. Hence, if the only $d_{\min}=2$
VNs are length-2 repetition codes, we have $P^{-1}(x) = \left( 1 /
\lambda_2^{(\rm rep)} \right) \, x$. This is the case for GLDPC
codes.
\end{property}

\medskip
\begin{property}\label{prop:2} For a length-$n_i$ SPC
CN (i.e. for a conventional LDPC degree-$n_i$ CN) we have $C_i =
n_i -1$.
\end{property}

\medskip
\begin{property}\label{prop:3}
Any length-$n_i$ and weight-2 binary sequence is a codeword for a
length-$n_i$ SPC CN. Then, $C_i$ for a length-$n_i$ CN with
minimum distance 2 is maximum when the CN is a SPC code. In other
words, $C_i$ fulfills
\begin{align*}
C_i & \leq n_i -1,
\end{align*}
\noindent where the equality holds when the CN is a SPC code.
\end{property}

\medskip
\begin{property}\label{prop:4}
For any VN with minimum distance $d_{\min}=2$, the value of
$P_i(x)$ depends on the chosen generator matrix through the
coefficients $A_{2,u}^{(i)}$. This is true for all $x$, except at
$x=0$ and at $x=1$ where we have
$$P_i(0) = 0$$
and
\begin{align*}
P_i(1) & = \frac{2\,A_2^{(i)}}{n_i} = C_i \\
\end{align*}
\noindent respectively. Independently of the VN representation,
the value assumed by $P_i(x)$ at $x=1$ is equal to the value of
$C_i$ for the same $d_{\min}=2$ linear block code when used as a
CN.
\end{property}

\medskip
\begin{property}\label{prop:5}
For a D-GLDPC code ensemble satisfying the derivative matching
condition, the iterative decoding threshold over the BEC assumes
the simple form $q^* = P^{-1}(1/C)$.

It should be noted that in general $P^{-1}(1/C)$ is not a closed
form for the threshold. However, there are simple cases in which
$P^{-1}(\cdot)$ can be explicited. An example is provided in the
appendix.
\end{property}


\section{Conclusion}\label{section:conclusion}
In this paper, a stability bound over the BEC has been developed
for D-GLDPC codes. It generalizes the inequality $q^* \leq
[\lambda'(0)\rho'(1)]^{-1}$, valid for LDPC code ensembles. We
have shown that for D-GLDPC codes, as for LDPC codes, the only
variable and check component codes contributing to the bound are
those having minimum distance 2. A derivative matching condition
sufficient to achieve the bound with equality has also been
defined. If the derivative matching condition is fulfilled, then
the decoding threshold over the BEC for D-GLDPC codes is expressed
by a simple formula, although in general not in closed-form. For
GLDPC codes this formula always leads to a closed-form threshold
expression.

\appendices
\section{D-GLDPC Codes with SPC Variable Nodes}
GLDPC codes employing strong generalized CNs (such as Hamming or
BCH CNs) represent a possible solution for obtaining a good
compromise between waterfall performance and error floor. Examples
of such GLDPC code constructions are described in
\cite{boutros99:generalized,lentmaier99:generalized,yue05:hadamard,tanner06:hybrid-journal,Liva2008:Tanner}.
In general, increasing the fraction of strong generalized CNs can
be very favorable from the point of view of the overall code
minimum distance and then of the error floor, but presents
drawbacks.

A first drawback is represented by an overall code rate loss which
makes GLDPC codes with large fractions of strong generalized CNs
of interest only for low or very low rate
\cite{fossorier05:generalized}. The reason is briefly reviewed
next. Let us consider a more general code structure, namely a
D-GLDPC code. If we denote by $r_{V,\, i}$ and by $r_{C,j}$ the
code rate of the type-$i$ VNs and of the type-$j$ CNs,
respectively, the overall design rate is
\begin{align}
R = 1 - \frac{\sum_j \, \rho_j \, (1 - r_{C,j})}{\sum_i \,
\lambda_i \, r_{V,\, i}},
\end{align}
which is monotonically increasing respect to any $r_{V,\, i}$ and
to any $r_{C,j}$. A generalized CN of length $n$ has a code rate
smaller than the code rate of a length-$n$ SPC CN. Then, a large
fraction of strong generalized CNs determines an overall rate
loss. In GLDPC codes this rate loss is difficult to compensate
even using large fractions of length-2 repetition VNs (which are
the highest rate VNs available if all the node in the Tanner graph
have minimum distance at least 2) so that usually the overall
GLDPC code remains of low rate.
A second drawback is that GLDPC codes with large fractions of
strong generalized CNs and large fractions of length-2 repetition
VN are typically characterized by a poor asymptotic threshold due
to the large area gap between the EXIT curves in the EXIT function
(see the Area Theorem in \cite{ten-brink04:design}).

Allowing the generalization of the VND together with the
generalization of the CND provides an increased flexibility in the
code design, that can be exploited to overcome the above mentioned
limitations. In particular, the rate loss due to the generalized
CNs can be compensated using generalized VNs with a code rate
larger than 1/2. In this context, a special class of generalized
VNs is represented by $(n,n-1)$ SPC VNs each one having $n$ edges
towards the CND and associated with $n-1$ encoded bits. It is
shown in \cite{wang07:DGLDPC_TCOM} that these codes can be
effectively exploited for the design of D-GLDPC codes with good
waterfall and error floor performance. In this appendix, we
develop the polynomial $P_i(\cdot)$ defined in Theorem
\ref{theorem:DGLDPC-bound} for such VNs when represented in both
systematic and cyclic form. We also propose a numerical example
illustrating the capabilities offered by D-GLDPC codes with SPC
VNs.

\subsection{SPC Variable Nodes in Systematic Form} Let us
suppose that the VNs of type-$i$ are length-$n_i$ SPC codes in
systematic form, i.e., represented by the $((n_i - 1) \times n_i)$
generator matrix
\begin{align*}
\mathbf{G}_i = \left[ \begin{array}{cccccc} 1 & 0 & 0 & \dots & 0 & 1\\
0 & 1 & 0 & \dots & 0 & 1\\
0 & 0 & 1 & \dots & 0 & 1\\
\vdots & \vdots & \vdots & \ddots & \vdots & \vdots\\
0 & 0 & 0 & \dots & 1 & 1\\
\end{array}\right].
\end{align*}
Each of these VNs has ${n_i \choose 2}$ weight-2 codewords.
Specifically, there are $n_i - 1$ weight-2 codewords generated by
weight-1 information words, ${n_i - 1 \choose 2}=\frac{(n_i -
1)(n_i - 2)}{2}$ weight-2 codewords generated by weight-2
information words and no weight-2 codewords generated by
information words of weight larger than 2. Then
$$
A_{2,u}^{(i)} = \left\{ \begin{array}{lcl} n_i - 1 & \textrm{if} &
u = 1\\
(n_i - 1)(n_i - 2)/2 & \textrm{if} & u = 2\\
0 & \textrm{if} & u = 3, \dots, n_i - 1
\end{array} \right.
$$
so that
\begin{align}\label{eq:SPC-VN-systematic}
P_i(x) & = \frac{2}{n_i} \cdot (n_i-1)\,\, x \, + \frac{2}{n_i}
\cdot
\frac{(n_i-1)(n_i-2)}{2} \,\, x^2 \notag \\
\, & = \frac{2(n_i - 1)}{n_i}\,\, x\, \left(1 + \frac{n_i -
2}{2}\,\, x \right).
\end{align}

\subsection{SPC Variable Nodes in Cyclic Form}

Let the VNs of type-$i$ be $(n_i, n_i-1)$ SPC codes in cyclic
form, i.e. generated by
\begin{align*}
\mathbf{G}_i = \left[ \begin{array}{cccccc} 1 & 1 & 0 & \dots & 0 & 0\\
0 & 1 & 1 & \dots & 0 & 0\\
0 & 0 & 1 & \dots & 0 & 0\\
\vdots & \vdots & \vdots & \ddots & \vdots & \vdots\\
0 & 0 & 0 & \dots & 1 & 1\\
\end{array}\right] \,\, .
\end{align*}
In this case we obtain an expression of $P_i(x)$ different from
\eqref{eq:SPC-VN-systematic}. In fact, it is readily shown that in
a SPC code represented in cyclic form, an information word of
weight $u$ generates a weight-2 codeword if and only if all its
`1' positions are contiguous. Then, for all $u = 1, \dots, n_i-1$
we have $A_{2,u}^{(i)} = n_i - u$, from which
\begin{align}\label{eq:SPC-VN-cyclic}
P_i(x) & = \sum_{u=1}^{n_i-1} \frac{2\,(n_i-u)}{n_i} \,\, x^u \notag \\
\, & = 2 \sum_{u=1}^{n_i-1} x^u - \frac{2}{n_i} \, \sum_{u=1}^{n_i-1} u \,\, x^u \notag \\
\, & = 2 \, x \, \frac{x^{n_i-1} - 1}{x - 1} - \frac{2\, x}{n_i}
\cdot \frac{1 - n_i \, x^{n_i-1} + (n_i-1) \, x^{n_i}}{(x - 1)^2} \notag \\
\, & = \frac{2\,x\,\left[ x^{n_i} - n_i \, (x - 1) - 1 \right]
}{n_i \,{(x - 1) }^2} \, \, .
\end{align}

\noindent If $n_i=2$ or $n_i=3$, then \eqref{eq:SPC-VN-systematic}
coincides with \eqref{eq:SPC-VN-cyclic} as expected. Specifically,
from both \eqref{eq:SPC-VN-systematic} and
\eqref{eq:SPC-VN-cyclic} we obtain $P_i(x) = x$ and  $P_i(x) =
\frac{2}{3}\, x^2 + \frac{4}{3}\, x$ for $n_i=2$ and $n_i=3$,
respectively.

\subsection{Comparison between Systematic and Cyclic Form}
Let us denote by $P_s(\cdot)$ and by $P_c(\cdot)$ the polynomial
$P_i(\cdot)$ of a length-$n$ SPC VN in systematic and cyclic form,
respectively. We show next that if $n
> 3$

$$
P_s(x) - P_c(x) \left\{ \begin{array}{lcl}
> 0 & \textrm{if} & 0 < x < 1 \\
= 0 & \textrm{if} & x = 1 \\
< 0 & \textrm{if} & x > 1 \, .
\end{array} \right.
$$

\noindent In fact, we have
\begin{align}
P_s(x) - P_c(x) & = \frac{2}{n}\, \left[ (n-1)\, x +
\frac{(n-1)(n-2)}{2} \, x^2 \right] - \frac{2}{n}\,
\sum_{u=1}^{n-1} (n-u)\, x^u \notag \\
\, & = \frac{2\, x^2}{n} \, \left[ \frac{(n-2)(n-3)}{2} -
\sum_{u=3}^{n-1} (n-u)\, x^{u-2} \right].
\end{align}
It is readily shown that $\sum_{u=3}^{n-1} (n-u) =
\frac{(n-2)(n-3)}{2}$. Then, $P_s(1) - P_c(1) = 0$, a result which
is consistent with Property \ref{prop:4}. For $0 < x < 1$ we must
have $\sum_{u=3}^{n-1} (n-u)\, x^{u-2} < \frac{(n-2)(n-3)}{2}$
which leads to $P_s(x) - P_c(x) > 0$; analogously, for $x > 1$ we
must have $\sum_{u=3}^{n-1} (n-u)\, x^{u-2} >
\frac{(n-2)(n-3)}{2}$ which leads to $P_s(x) - P_c(x) < 0$.

\subsection{D-GLDPC Codes with Length-2 Repetition VNs and SPC VNs in Systematic Form}

Let us consider \eqref{eq:general-stability-cond}. Although in
general it is not possible to express $P^{-1}(\cdot)$ in an
explicit closed form, this is possible in special cases. For
instance, obtaining a closed form expression of $P^{-1}(\cdot)$ is
possible when the only $d_{\min} = 2$ variable component codes are
length-2 repetition codes and length-$n$ SPC codes in systematic
form. Let $\lambda$ be the fraction of edges connected to the
length-2 repetition VNs and $\mu$ the fraction of edges connected
to the length-$n$ SPC VNs (so $\lambda+\mu$ is the total fraction
of edges connected to $d_{\min}=2$ VNs). We have
\begin{align*}
P(x) & = \lambda \, x + \mu \, \frac{2(n - 1)}{n}\,\, x\,
\left(\frac{n - 2}{2}\,\, x + 1 \right).
\end{align*}
\noindent By solving for positive $y$ the equation $P(y) = x$, we
obtain

\begin{align}\label{eq:P^(-1)-mixture}
P^{-1}(x) = \frac{-\left[ n\,\lambda  + 2\,\left( n-1 \right)
\,\mu  \right] }{2\,\left( n-2 \right) \,\left( n-1 \right) \,\mu
} + \frac{{\sqrt{{\left[ n\,\lambda  + 2\,\left( n-1 \right) \,\mu
\right] }^2 + 4\,\left( n-2 \right) \,\left( n-1 \right) \, n\,
\mu\, x }}}{2\, \left( n-2 \right) \,\left( n-1 \right) \,\mu } \,
\, .
\end{align}

\bigskip

\begin{figure}[!t]
\psfrag{x}{$x$} \psfrag{P-1(x)}{$P^{-1}(x)$}
\begin{center}
\includegraphics[width=10.5 cm, angle=270]{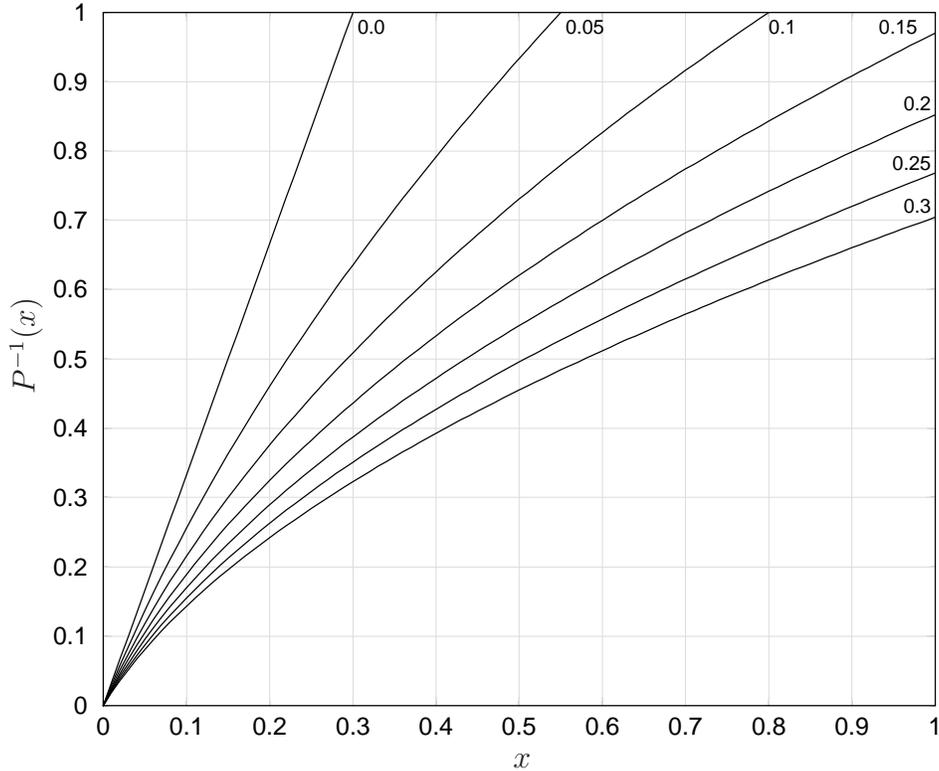}
\end{center}
\caption{Plot of $P^{-1}(\cdot)$ for a D-GLDPC code where the only
$d_{\min}=2$ VNs are length-2 repetition and length-7 SPC VNs. The
total fraction of edges connected to $d_{\min}=2$ VNs is
$\lambda+\mu=0.3$, and each curve is associated with a specific
value of $\mu$.} \label{fig:P^(-1)-mixture}
\end{figure}

In Fig. \ref{fig:P^(-1)-mixture}, %
\eqref{eq:P^(-1)-mixture} is plotted for different values of
$\mu$, assuming $\lambda + \mu = 0.3$ and SPC VNs of length $n=7$.
Each curve is associated with a different value of $\mu$, i.e.,
with a different proportion between length-2 repetition VNs and
length-7 SPC VNs in the VND. Hence, the curve labelled as $0.0$
corresponds to the presence of only length-2 repetition VNs, while
the curve labelled as $0.3$ to the presence of only SPC VNs. Hence
modifying $\mu$ provides a wide variety of options.

\subsection{Distribution Optimization}
\begin{figure}[t]
\begin{center}
\psfrag{IAVIEC}[cc]{$I_A$} \psfrag{IEVIAC}[cc]{$I_E$}
\psfrag{IEV}[cc]{$I_{E,V}(I_A,q^*)$}
\psfrag{IEC}[cc]{$I_{E,C}^{-1}(I_A)$}
\includegraphics[width=10.5 cm, angle=270]{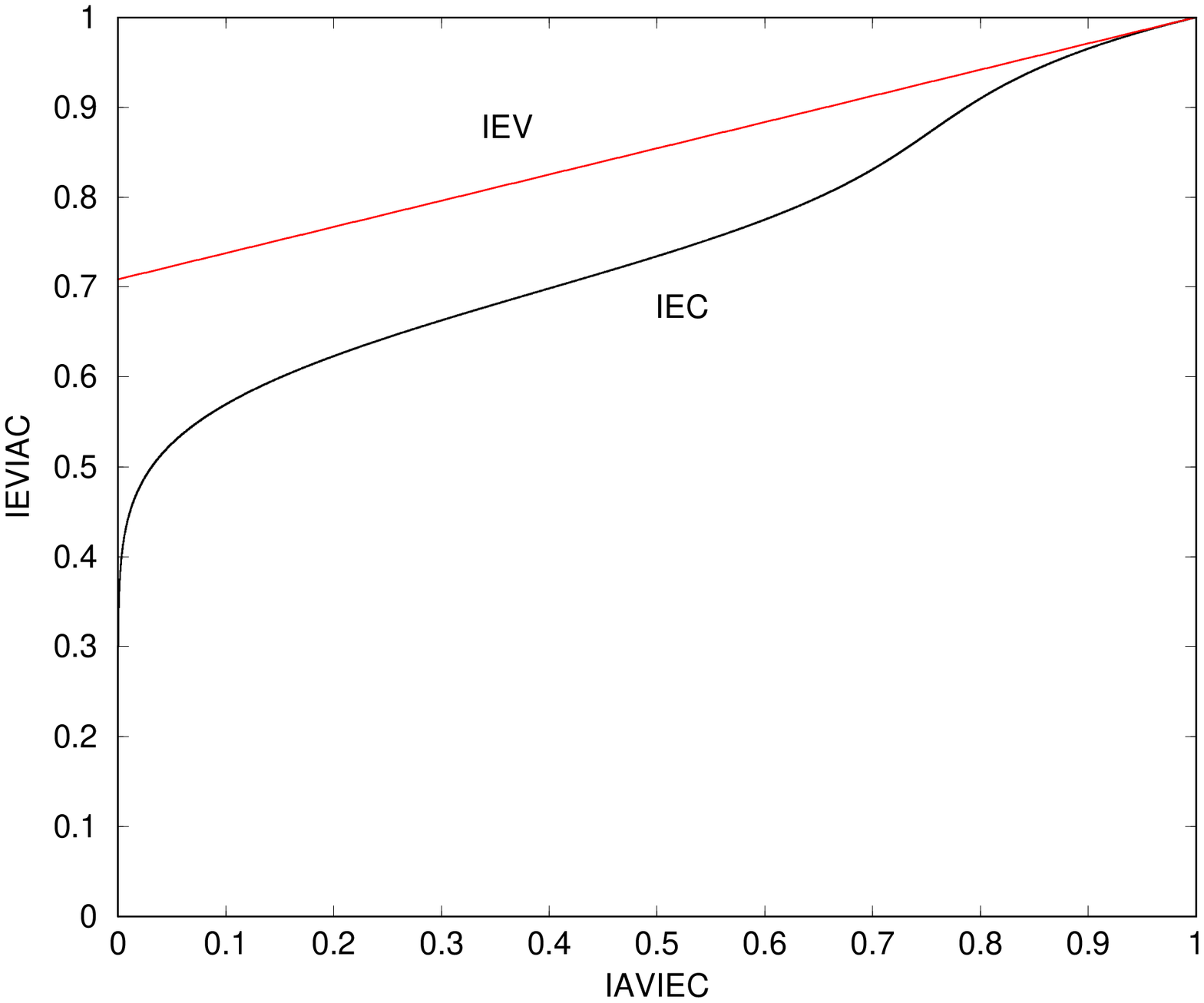}
\end{center}
\caption{EXIT chart for the GLDPC distribution in Table I.}
\label{fig:EXIT-BCH-rep2}
\end{figure}

\begin{figure}[t]
\begin{center}
\psfrag{IAVIEC}[cc]{$I_A$} \psfrag{IEVIAC}[cc]{$I_E$}
\psfrag{IEV}[c]{$I_{E,V}(I_A,q^*)$}
\psfrag{IEC}[cc]{$I_{E,C}^{-1}(I_A)$}
\includegraphics[width=10.5 cm, angle=270]{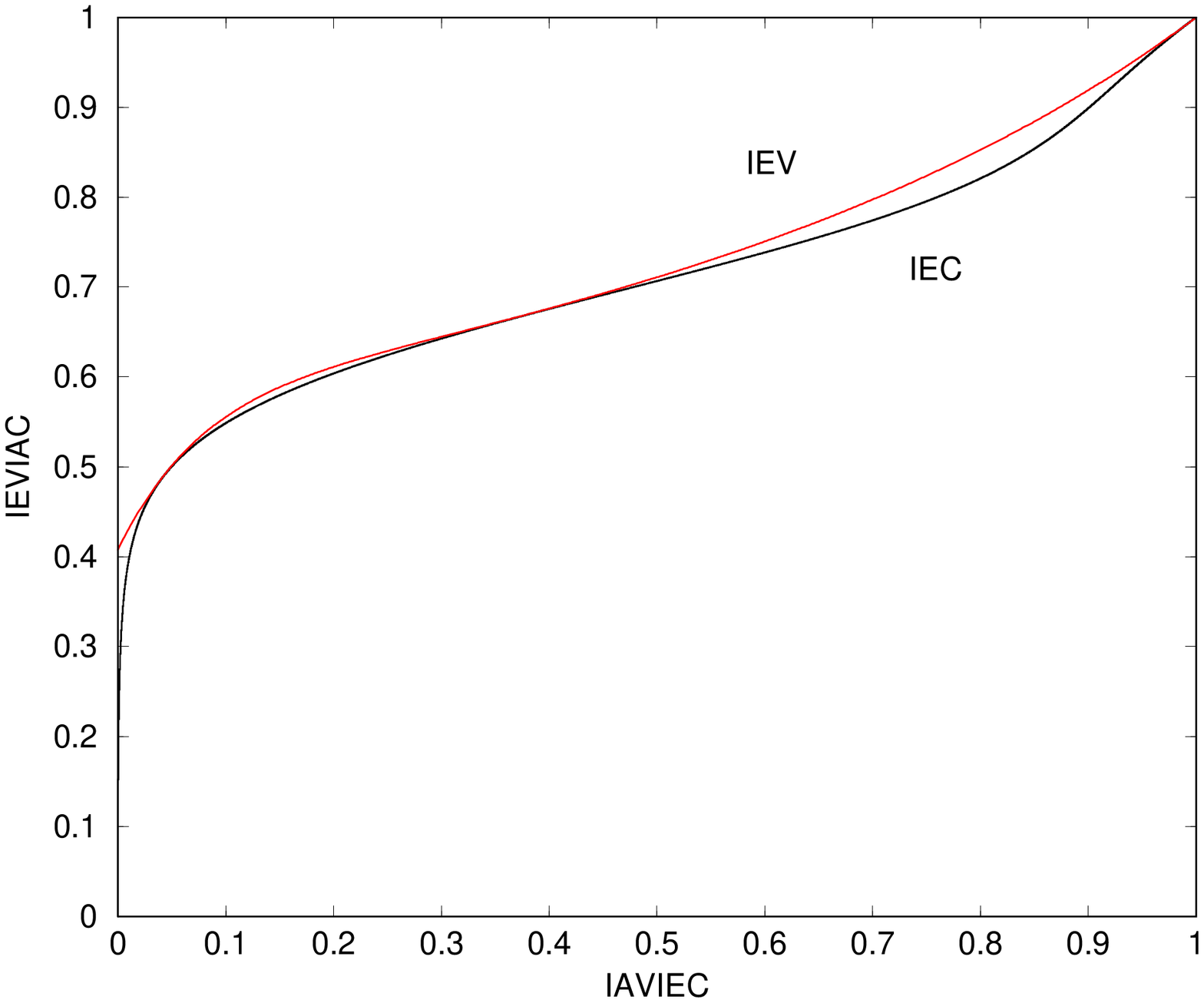}
\end{center}
\caption{EXIT chart for the D-GLDPC distribution in Table I.}
\label{fig:EXIT-BCH-SPC15}
\end{figure}

We consider the optimization problem of a GLDPC and of a D-GLDPC
code ensemble for design rate $R=1/2$. In both cases we constrain
the optimization process by allowing the repetition VN degree to
range only between 2 and 15 and the SPC CN degree only between 5
and 15. Moreover, we use $(31,21)$ BCH CNs, imposing a minimum
fraction of edges connected to the BCH CNs equal to 0.7. For the
D-GLDPC code ensemble, we allow also length-15 SPC CNs in cyclic
form. The output of an optimization process over the BEC performed
with differential evolution
\cite{price97:differential-evolution,shokrollahi00:design} is
reported in Table I (from an edge perspective). For each of the
two optimized distributions the threshold and the stability bound
\eqref{eq:general-stability-cond} are shown. While for the GLDPC
code ensemble it is necessary to use only length-2 repetition VNs
to compensate the rate loss introduced by the large fraction of
BCH CNs with an overall poor threshold, for the D-GLDPC code
ensemble the use of SPC VNs allows obtaining a much larger
threshold. From an EXIT chart perspective the capability of the
SPC VNs to reduce the area gap between the EXIT curves is
illustrated by comparing in Fig. \ref{fig:EXIT-BCH-rep2} and Fig.
\ref{fig:EXIT-BCH-SPC15}.

\begin{table}\label{table:strong-CND-weak-VND}
\caption{GLDPC and D-GLDPC distributions with large fractions of
BCH check nodes}
\begin{center}
\begin{tabular}{|l|c|c|c|}
\hline
&  GLDPC & D-GLDPC\\
\hline\hline
\multicolumn{3}{|c|}{\emph{Variable Nodes}}\\
\hline \hline
SPC$_\textrm{cyc}$ 15 &  & 0.521581\\
rep 2 & 1.000000 &  0.132836\\
rep 14 &  &  0.145293\\
rep 15 &  & 0.200291\\
\hline \hline
\multicolumn{3}{|c|}{\emph{Check Nodes}}\\
\hline \hline
BCH & 0.700000  & 0.721799\\
SPC 5 &   & 0.278201\\
SPC 12 & 0.174190 &  \\
SPC 13 & 0.125810 &  \\
\hline \hline
$q^*$ & 0.291516 & 0.478585\\
\hline
$P^{-1}(1/C)$ & 0.291902 & 0.478585\\
\hline
\end{tabular}
\end{center}
\end{table}
%

%
%
%
%
\bibliography{IEEEabrv,D:/epaolini/works/Bib/bibfile}
\end{document}